\documentclass[pdflatex,sn-mathphys-num]{sn-jnl}

\usepackage{graphicx}%
\usepackage{multirow}%
\usepackage{amsmath,amssymb,amsfonts}%
\usepackage{amsthm}%
\usepackage{mathrsfs}%
\usepackage[title]{appendix}%
\usepackage{xcolor}%
\usepackage{textcomp}%
\usepackage{manyfoot}%
\usepackage{booktabs}%
\usepackage{algorithm}%
\usepackage{algorithmicx}%
\usepackage{algpseudocode}%
\usepackage{listings}%

\theoremstyle{thmstyleone}%
\theoremstyle{thmstyletwo}%

\theoremstyle{thmstylethree}%

\begin{document}

\title[Light-induced Non-uniform Surface Polarization Evolution as Active Matter Motion Mechanism]{Light-induced Non-uniform Surface Polarization Evolution as Active Matter Motion Mechanism}


\author*[1]{\fnm{Bohdan} \sur{Lev}}\email{bohdan.lev@gmail.com}

\author[1]{\fnm{Oleksandr} \sur{Cherniak}}\email{anchernyak@bitp.kyiv.ua}

\affil*[1]{\orgdiv{Synergetics}, \orgname{Bogolyubov Institute for Theoretical Physics of NAS of Ukraine}, \orgaddress{\street{Metrolohichna 14-b}, \city{Kyiv}, \postcode{03143}, \state{Kyiv}, \country{Ukraine}}}




\abstract{A new mechanism for micro-swimmer motion in an aquatic environment, which implies light-induced polarization perturbations as a motivation for its motion, is proposed. The mechanism assumes that natural light leads to an enhancement of the non-uniform polarization distribution at the micro-swimmer's surface and also to an enhancement of its surface deformations. The interaction of such non-uniform polarization with free charges of the environment and the mechanical interaction of surface deformations with media flows lead to the directed micro-swimmer motion. The mathematical justification for this mechanism is based on the free energy functional minimization in terms of polarization. The obtained solutions correspond to the translation motion along the micro-swimmer and the rotation of the non-uniform polarization distribution. The estimation of the velocity of micro-swimmer motion, which is based on the hydrodynamics of the active matter, corresponds to the experimentally observed values.}




\maketitle

\section{Introduction}\label{sec1}

There are a variety of motion mechanisms \cite{Yad} for different microscopic swimmers, like mechanical \cite{Dre, Knut}, chemical \cite{Jang2023}, thermal \cite{Alo, Pav}, or other processes \cite{Sur, Bus, Ezh, Sto, Pon, Hun}. The two most common mechanisms for the living micro-swimmers use special structures, i.e., flagellum or fimbria, and their manipulations, like rotary motion of the bacterium's long offshoot that is connected to the bacterium's surface by one end or twitches of special hair-like formation on the bacteria's surfaces, to provide swimming or crawling mobility correspondingly. However, mentioned structures appear to be energy-demanding ones \cite{DiNezio2024} and therefore for artificial active matter \cite{pnas2500526122, smll202501317, adu8009} or smaller structures like DNA \cite{Shi2024, Shim2024} another mechanisms, which particularly rely on external electric fields, are considered. It is also worth mentioning the motile bacteria without structures like a flagellum or a fimbria, whose motility is provided by a gliding mechanism, like Myxococcus Xantus, which employs a variable set of diverse motor complexes like focal adhesive ones \cite{Mauriello2010, Chen2023, Grandy2023, Cambillau2025}, or Spiropasma, which has a unique helical-shaped elastic cytoskeleton \cite{WOLGEMUTH2003828, Lopez20, Wada09, Yang09}. Such bacteria, without flagellum or fimbria, move much more slowly than ones which poses these structures, typically by one or two magnitudes, that approximatly correspond to the artificial micro-swimmer's velocity.

In this work, we propose a motion mechanism for active matter that is based on light-induced self-consistent polarization redistribution at the micro-swimmer surface. There are a few conditions that need to be met for the realization of this mechanism. The first one is the ability of ambient light to develop or enhance non-uniform polarization at the surface, which means some kind of phototactic properties for the micro-swimmer. The next one is the elasticity of the surface to emerge the structurization on the surface governed by such non-uniform surface polarization. And the last one is the existence of free charged particles in media to interact with a non-uniform polarization distribution. Based on these conditions, we formulate the active matter motion mechanism as follows. The ambient light intensifies polarization redistribution both within the micro-swimmer volume and on its surface, and leads to the development or enhancement of the non-uniform polarization distribution on the micro-swimmer surface. A momentum for the whole micro-swimmer is provided by the evolution of such a non-uniform polarization distribution by two mechanisms. The first one, named polarization mechanism, is governed by electrical interaction between non-uniform surface polarization distribution and charged particles in the aqueous medium \cite{Jarisz2018, Ngo2021, Phan2022, pnas2500526122, smll202501317, adu8009, Kilic2011, Boymelgreen2016, Katzmeier2023, OHSHIMA1991191}. The second one, the deformation mechanism, is provided by the mechanical interaction between the bacterium's surface deformations and the medium's flows \cite{Taylor1951, Lighthill1952, Palagi2016, Gelebart2017}. We demonstrate that among all possible non-uniform polarization distributions and surface deformations, there are ones that can generate non-equivalent distributions and deformations that provide directional motion of the bacterium in accordance with the scallop theorem \cite{Purcell, C0SM00953A}. These light-induced non-equilibrium structures, i.e., non-uniform polarization distribution and surface deformations, vanish as soon as the source that causes them, i.e., ambient light, vanishes. We consider such structures or patterns as persistent heterogeneities (solitons) limited by the bacterium's surface. They evolve along the surface towards the symmetry-violated domain, as it has been shown earlier both analytically and in terms of numerical simulation \cite{HYMAN198123}. The velocity propagation \cite{jcp3158471, PhysRevE.56.3265} estimations of such solitons are considered in a similar way as excited molecules concentration wave propagation, i.e., pattern propagation, with specified velocity along Langmuir monolayer as demonstrated in \cite{Tabe2003, Rebesh2022}.

We use a thermodynamic approach \cite{Seul, Matt}, namely free energy functional minimization procedure, to demonstrate the possibility of generating such non-uniform polarization distributions and surface deformations that are able to provide a sufficient impulse for the micro-swimmer to have a directed motion in principle. We estimate micro-swimmer's velocity based on the active matter hydrodynamics equation \cite{Mar, Wen, Sid, Kos2020, KosRavnik22}. The partial justification of the assumptions made for this mechanism and some aspects of the ambient light effect on a micro-swimmer, in particular, the living one, in an aqueous medium, is discussed in the section Discussion. The details of ambient light effects on micro-swimmer surface or various processes inside the micro-swimmer's surface remain beyond the framework of this paper, in particular, for the sake of the mechanism's versatility. We only note that several works that study in detail such processes separately as: change in the refractive index due to polarization with details on the propagation speed of the concentration wave of excited molecules \cite{Tabe2003, Rebesh2022}, laser manipulation of an object \cite{Iti, Kal, Uts, Mark} or manipulation of individual molecules during thermophoresis \cite{Bra}, surface changes \cite{Peterson2021, Miele2022}, pattern formation \cite{Sam}, generation of complex structures \cite{Dog, Han, Wang} for different specific systems.

So, in this paper, we propose a new hypothetical mechanism for a micro-swimmer motion in an aquatic environment that relies on light-induced polarization perturbations at the surface of the micro-swimmer and assumes that mechanical deformations in the presence of bound degrees of freedom inside and outside the micro-swimmer can lead to its translational motion. The aim of this work is to introduce the basic concepts of this mechanism and stimulate experimental studies. We do not describe all of the peculiarities of energy conversion of chemical or other processes inside the active matter that change polarization, but rather assume where this is needed. Finally, some of them we partially address in the Discussion section.

\section{Results}\label{sec2}

Here, we present the mechanism of light-induced non-equilibrium structures formation and the equations of evolution thereof in terms of the free energy functional. We start with the formation of non-uniform polarization at the micro-swimmer's surface of oblong drop-like shape, which we approximate as a cylinder or an ellipsoid. After the demonstration of the possibility of such polarization distribution formation, we consider the immediate side-effect - surface deformations. When desired non-equivalent structures are present, we estimate the micro-swimmer?s velocity based on active matter hydrodynamics equations.

\subsection{Polarization mechanism}
Let us start with the free energy analysis of the active matter proposed in the review Ref.\cite{Mar}. Since an arbitrary non-uniform polarization distribution on the surface of a micro-swimmer is a non-equilibrium structure, we need to take into account the non-linear terms in our free energy functional, which is provided by the feedback function. A natural way is to assume that the elasticity coefficient depends on polarization in the free energy functional, similar to Ref.\cite{Mar}, which is given by
\begin{eqnarray}
	F=&\int dV \left( f + \delta f \right) = \int dV \left(\frac{\alpha}{2}|\mathbf{P}|^{2} + \frac{\beta}{4}|\mathbf{P}|^{4} + \right.\nonumber\\ &\left.\left( K + \kappa |\mathbf{P}|^2 \right) \left[ (\nabla \cdot \textbf{P} )^2 + (\nabla \times \textbf{P} )^ 2 \right] \right). \label{eqLevPolFreeEnergyP2}
\end{eqnarray}
It provides the desired feedback and is physically substantiated with continuous mean-field order-disorder parameters $\alpha$, $\beta$ and Frank constant $K$ that describes the free-energy increase due to the distortion of the ordered configuration. It should be noted right away that in this continuum model Ref.\cite{Mar}, which we also use later in this work, the polarization vector plays a dual role simultaneously: as a parameter of the orientational order and as a velocity field of a separate surface region $\upsilon_{\textbf{P}}(\textbf{r})$. We approximate the oblong shape of the micro-swimmer drop by a cylinder and polarized globules within the micro-swimmer by polarized spheres. Thermophoresis-intensified motion of polarized globules leads to the formation of an inhomogeneous polarization distribution. We assume that polarization within the micro-swimmer is zero, and hence, inhomogeneous polarization distribution occurs only at the micro-swimmer's interface. We also assume that polarization distribution evolves faster than a rearrangement of free charges in the aqueous medium that might screen it. This guarantees the absence of free charges on the micro-swimmer's surface, so we have $4 \pi \rho = \nabla \cdot \mathbf{D} = \nabla \cdot \left( \mathbf{E} + 4 \pi \mathbf{P} \right) = \nabla \cdot \mathbf{P} = 0,~r = r_{S}$ from Maxwell's equation. Finally, since we assume that the torque for polarized globules is nonzero in the case of the surface deformation mechanism, it is reasonable to calculate the rotary inhomogeneous polarization distribution along the micro-swimmer's cylinder axis, thus we have $ \left( \nabla \times \mathbf{P} \right)^2 = \omega^ 2\left( \nabla |\mathbf{P}| \right)^2$. In terms of the above assumptions, employing the Euler-Lagrange equation, we obtain the equation for the inhomogeneous polarization distribution given by
\begin{equation}
	\phi + c_{\beta} \phi^3 - c_{\kappa} \phi \dot{\phi}^2 - \left( K_{\alpha} + c_\kappa \phi^2  \right) \ddot{\phi} = 0, \label{eqLevPolFreeEnergyPhiMinDetail}
\end{equation}
for new convenient variables $|\mathbf{P}| = \phi$, $|\nabla |\mathbf{P}|| = \dot{\phi}$, $c_{\beta} = \beta / \alpha$, $K_{\alpha} = 2 K \omega^2 / \alpha$ and $c_{\kappa} = 2 \kappa \omega^2 / \alpha$. It is impossible to find the solution of Eq.~\ref{eqLevPolFreeEnergyPhiMinDetail} analytically, so we consider the approximations given below for various relations between the perturbation $\kappa |\mathbf{P}|^2$ and Frank constant $K$.

\begin{figure*}[htp!]
	\includegraphics[width=0.49\textwidth]{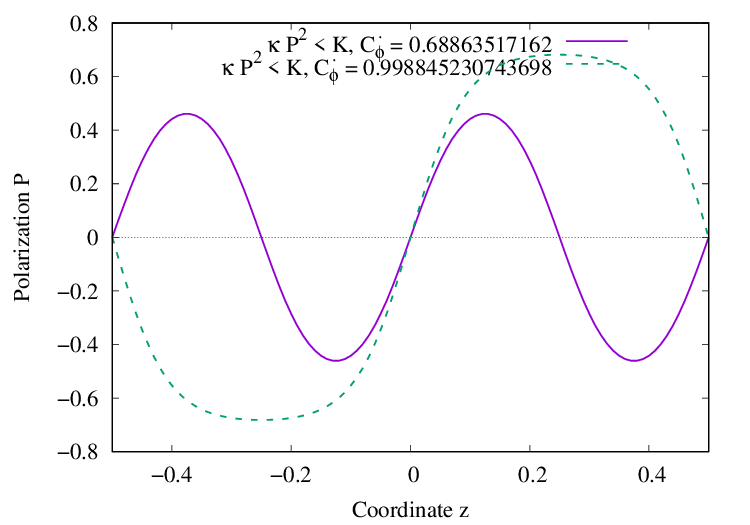} 	\includegraphics[width=0.49\textwidth]{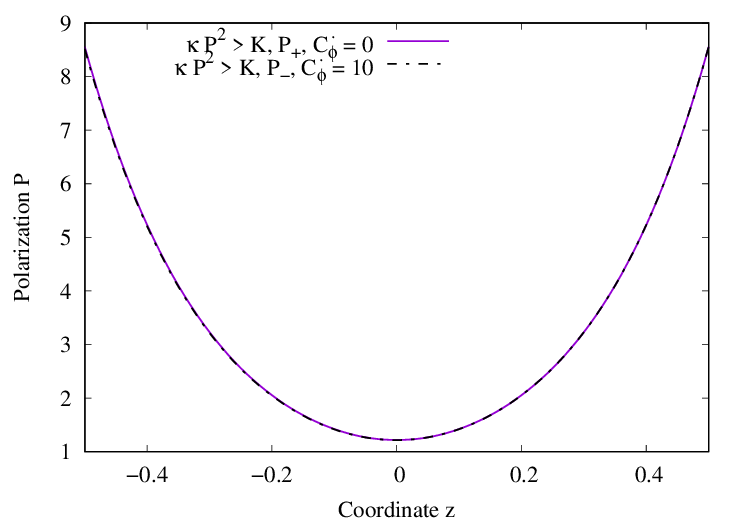}
	\includegraphics[width=0.49\textwidth]{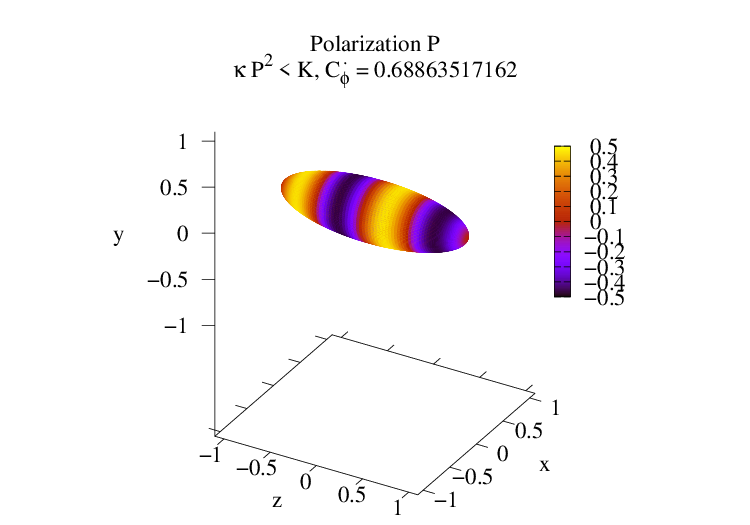} 	\includegraphics[width=0.49\textwidth]{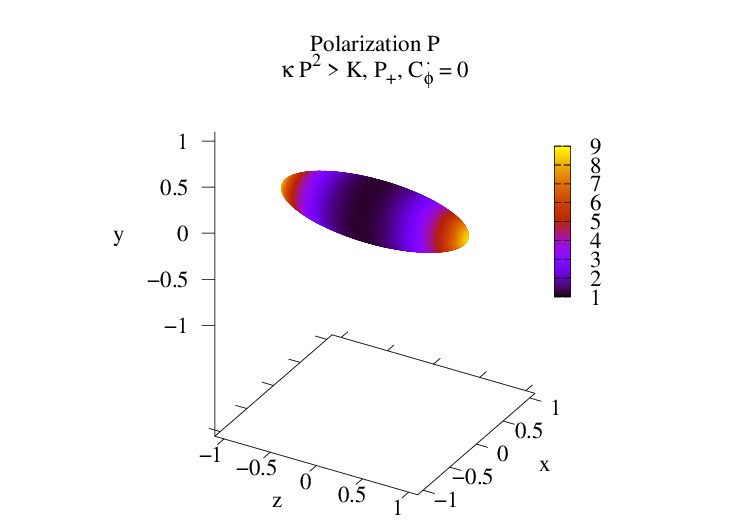}
	\includegraphics[width=0.49\textwidth]{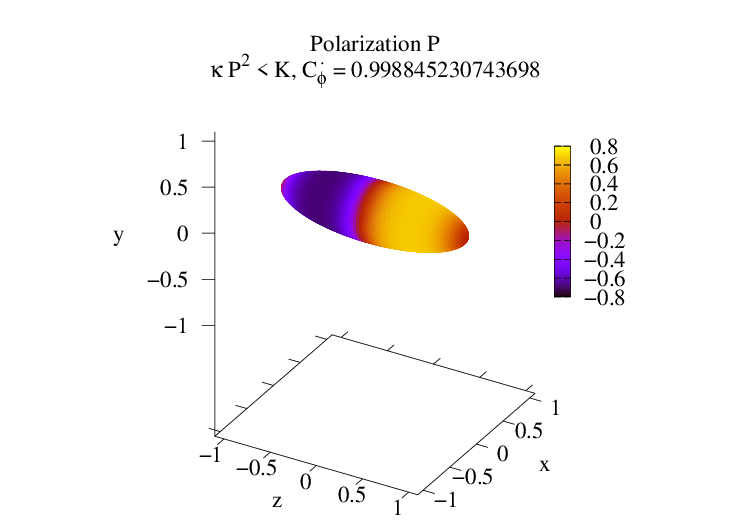} 	\includegraphics[width=0.49\textwidth]{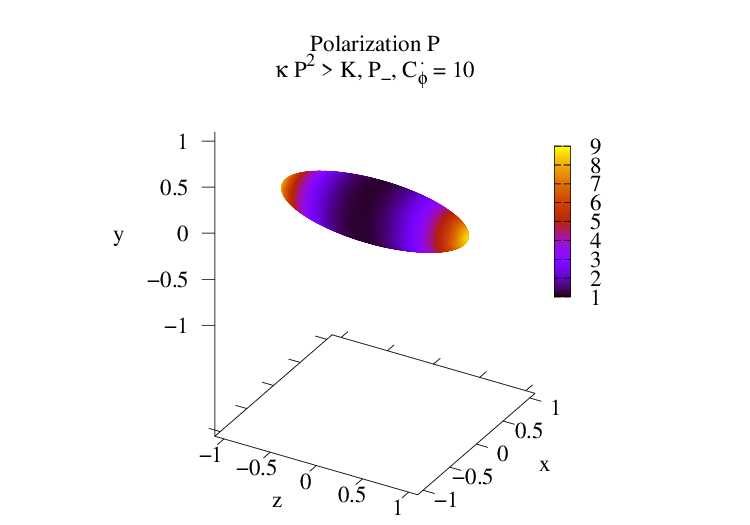}
	\caption{Polarization profiles on the micro-swimmer's interface. On the left -- for solutions for the polarization with zero perturbation and $\phi(\pm l/2) = 0$ at the ends of a micro-swimmer of length $l$. On the right, in the case of solutions for polarization with strong perturbations. The coordinate $\mathrm{z}$ is measured in terms of the micro-swimmer's length $l \approx 2 \times 10^{-3} \mathrm{cm}$. The middle and bottom left figures demonstrate polarization magnitude distribution for solutions for the small perturbations presented in the top left figure. The middle and bottom right figures demonstrate polarization magnitude distribution for solutions for the large perturbations presented in the top right figure.}
	\label{fig_pKappaSmall:fig1}
\end{figure*}

If the perturbation $\kappa |\mathbf{P}|^2$ is negligibly small, i.e., $c_{\kappa} / K_{\alpha} \rightarrow 0$, we rewrite Eq. \ref{eqLevPolFreeEnergyPhiMinDetail} as given by
\begin{equation}
	\phi + c_{\beta} \phi^3 - K_{\alpha} \ddot{\phi} = 0, \label{eqLevPolFreeEnergyPhiKappaSmall}
\end{equation}
This equation may be regarded as Painleve-VIII, its solution is known to be given by the elliptic Jacobi function
\begin{eqnarray}
	\phi = \frac{ \sqrt{1 + \sqrt{ 1 - 2 C_1 K_{\alpha} c_{\beta} } } }{ \sqrt{ - c_{\beta} } } \times\nonumber\\\mathrm{Sn} \left( \sqrt{ \frac{ - c_{\beta} C_1 \left( C_2 + z \right)^2 }{ 1 + \sqrt{ 1 - 2 C_1 K_{\alpha} c_{\beta} } } } , \frac{ 1 + \sqrt{ 1 - 2 C_1 K_{\alpha} c_{\beta} } }{ 1 - \sqrt{ 1 - 2 C_1 K_{\alpha} c_{\beta} } } \right), \label{eqLevPhiSolKappa0_C2C1}
\end{eqnarray}
where both constants $C_2$ and $C_1$ are determined by the initial conditions for polarization and its gradient. Inasmuch as the double-periodic Jacobi function $\mathrm{Sn}(z, k)$ is odd and micro-swimmer's ends are considered in $\pm l/2$, we put the polarization at micro-swimmer's ends to be equal to zero, $C_2 = \phi\left( 0 \right) = 0$, and the polarization gradient to be given by $C_1 = \left( \dot{\phi}^2(0) - \frac{{\phi}^2(0)}{K_{\alpha}} - \frac{c_{\beta}{\phi}^4(0)}{2 K_{\alpha}} \right) = C_{ \dot{\phi} }^2$. We treat the latter as a free parameter $C_{ \dot{\phi} }$, however, limited by condition $C_{ \dot{\phi} } <  1/\sqrt{2 K_{\alpha} c_{\beta}} $ to obtain real values for polarization. So, we put $C_{ \dot{\phi} } <  1/\sqrt{2 K_{\alpha} c_{\beta}} $ and polarization at the micro-swimmer's ends to be equal to zero. Polarization profile for the two specified values of $C_{ \dot{\phi} }$ is shown in Fig.~\ref{fig_pKappaSmall:fig1} (Left).

The interpretation of the solutions thus obtained is given in what follows. Moderate perturbations of the polarization distribution produce an inhomogeneous polarization profile on the micro-swimmer's surface. This distribution rotates about the micro-swimmer's cylinder axis faster than charged particles in the water are localized on the micro-swimmer's surface and screen its polarization. The evolution of the nonuniform polarization distribution interacts via the electric force with the surrounding charges and causes cumulative displacement of the micro-swimmer. The micro-swimmer seems to be screwed in the surrounding medium.

For sufficiently large values of the perturbation, $\kappa |\mathbf{P}|^2 \gg K$, i.e., for  $K_{\alpha}/c_{\kappa} \rightarrow 0$, the term containing $K_{\alpha}$ may be disregarded, so Eq.\ref{eqLevPolFreeEnergyPhiMinDetail} is reduced to the form  
\begin{equation}
	1 + c_{\beta} \phi^2 - c_{\kappa} \dot{\phi}^2 - c_\kappa \phi \ddot{\phi} = 0, \label{eqLevPolFreeEnergyK0}
\end{equation}
The latter equation may be transformed to the Bernoulli equation by introducing a new variable $u(\phi) = u_1(\phi) u_2(\phi) = \dot{\phi}^2$.
\begin{eqnarray}
	u(\phi) = \frac{c_u}{\phi^2} + \frac{1}{c_{\kappa}} + \frac{c_{\beta}}{2c_{\kappa}} \phi^2 = \dot{ \phi}^2 \label{eqLevUSol}
\end{eqnarray}
with the new constant $c_u = \left(2 c_{\kappa} C_1^2 \phi_0^2 - 2 C_2^2 - c_{\beta} C_2^4 \right)/ 2 c_{\kappa}$. We integrate Eq.\ref{eqLevUSol} to obtain $z(\phi)$, thus, we find the polarization for the case of strong perturbations to be given by
\begin{eqnarray}
	\phi =\pm \left(\frac{ \exp{ \left( \sqrt{ \frac{2 c_{\beta}}{c_{\kappa}}} z \right) }}{ 2 }  - \frac{1}{c_{\beta}} - \right.\nonumber\\ \left.\frac{\left( \frac{2 c_u c_{\kappa}}{c_{\beta}} - \frac{1}{c_{\beta}^2} \right) \exp{ \left( - \sqrt{ \frac{2 c_{\beta}}{c_{\kappa}}} z \right) } }{ 2 } \right)^{1/2}. \label{eqLevPolSol}
\end{eqnarray}
where $C_2 = \phi_0 = 1.21708, 1.21909$ and $C_1 = \dot{\phi}_0= C_{\dot{\phi}} = 0, 10$ are free parameters. The relevant solutions are given in Fig.~\ref{fig_pKappaSmall:fig1} (Right).

In the case of strong polarization perturbations, the maxima of the polarization distribution are localized at the micro-swimmer's ends. Hence, we assume that the induced repulsion might be the reason for the micro-swimmer's fragmentation. Indeed, the polarization of some signs is localized at the ends of the micro-swimmer and thus causes stretching of the ends and probable fragmentation of the micro-swimmer.

\subsection{Deformation mechanism}

The resulting inhomogeneous polarization distribution is associated with a change in the concentration and orientation of dipole moments that generates deformations of the microswimmer surface. To understand this process, we consider the distribution of molecular dipoles proposed in Ref.\cite{Seul}. Their moments are directed perpendicular to the surface and are involved in pairwise electrostatic dipole-dipole interaction. Since we aim to find such surface deformations that can lead to micro-swimmer directional motion, we consider only local ones and just assume there is a negligibly small polarization in the bulk volume. In the continuous approximation, the dipole-dipole interaction is given by an integral over a two-dimensional monolayer with the order parameter $c(\mathbf{r})$ that is the concentration of dipoles of polarized regions near the surface, i.e.,
\begin{equation}
	F_{d}=\frac{1}{2} \int dS \left( b (\mathbf{\nabla} c(\mathbf{r}))^{2}+f(\left(\left|c( \mathbf{r})\right|\right)^{2}-1)^{2} \right) \label{1}
\end{equation}
In the case of small variations of the concentration, the term with quadratic gradient and "stiffness" coefficient $b$ is given rise to by energy expenditure for surface formation in the lowest approximation order. Dipole distribution depends on the physical processes associated with the luminosity intensity $I$ and the value of the temperature gradient $\mathbf{\nabla }T$, i.e., $b = b(I,\mathbf{\nabla}T)$. To study the nonlinear effects, we need to consider the coefficient $b = b(I, c(\mathbf{r}),\mathbf{\nabla}T)$ that depends on the dipole concentration, and its importance will be discussed later. Instead of considering the far interaction of molecular dipoles, we study the deformation and bending of the surface similarly to Ref.\cite{Seul}. So, we write the interaction between the height profile relative to the flat state $h(\mathbf{r})$ and the local concentration $c(\mathbf{r})$ of polarized dipoles as given by
\begin{eqnarray}
	F_{s}=&\frac{1}{2}\int dS \left( \sigma \left( \left| \mathbf{\nabla} h(\mathbf{r}) \right| \right)^{2} + k \left( \left| \mathbf{\nabla^{2}} h(\mathbf{r}) \right| \right)^{2} + \right.\nonumber\\ &\left.\lambda c(\mathbf{r}) \left( \mathbf{\nabla^{2}} h(\mathbf{r}) \right) \right)& \label{2}
\end{eqnarray}
Which is one among the possible forms to describe the deformation and bend of the surface, where $\sigma$ is the surface tension for the flat surface, $k$ is its bend module, and the coefficient $\lambda$ in the last term describes the coupling of the local curvature $\mathbf{\nabla^{2}} h(\mathbf{r})$ and local concentration $c(\mathbf{r})$. We note that the latter term in Eq.~\ref{2} relates the local curvature $\mathbf{\nabla^{2}} h(\mathbf{r})$ and the local dipole concentration $c(\mathbf{r})$, when the bend module $k$ depends on the local dipole concentration. In this approach, we are not interested in the shape of the micro-swimmer, which is associated with the mean curvature, and the direction of polarization, which is associated with the divergence of the surface normal. The sum of free energies  $F=F_{d}+F_{s}$ completely describes the equilibrium state in terms of the dipole distribution on the interface between the micro-swimmer and aqueous medium. It should be noted that the term with coefficient $f=0$  of Eq.~\ref{1} may be disregarded since dipole-dipole interaction can induce only orientation ordering rather than spatial ordering. This implies that the term that takes into account the nonlinearity cannot be determined by the dipole distribution. Besides, since dipole distribution within the volume of the micro-swimmer determines the curvature of its surface, and the surface is closed by an oblong drop, the formation of a nonuniform dipole distribution produces an inhomogeneous corrugated surface that can move. Thus, the problem is reduced to the consideration of the conditions of surface deformation. The relation for the dipole concentration gradient and surface curvature directly follows from the minimum of the sum of free energies with respect to the height profile $h(\mathbf{r})$ and dipole concentration $c(\mathbf{r})$. The Euler-Lagrange equation is reduced to the equation for the concentration
\begin{equation} \label{4}
	b\mathbf{\nabla}^{2} c(\mathbf{r})-\lambda\mathbf{\nabla}^{2}h(\mathbf{r})=0
\end{equation}
and the equation for the surface profile
\begin{equation} \label{6}
	-\sigma \mathbf{\nabla^{2}}h(\mathbf{r})+k\mathbf{\nabla^{2}}\mathbf{\nabla^{2}} h(\mathbf{r} )+\lambda\mathbf{\nabla^{2}} c(\mathbf{r})=0
\end{equation}
Thus, we can write Eq.~\ref{6} in terms of curvature only, i.e.,
\begin{equation}
	k\mathbf{\nabla^{2}}\mathbf{\nabla^{2}} h(\mathbf{r})=(\sigma-\frac{\lambda^{2} }{b})\mathbf {\nabla^{2}} h(\mathbf{r}) \label{5}
\end{equation}
The latter is the Helmholtz equation, whose general solution for the two-dimensional Euclidean space may be written as given by
\begin{eqnarray}
	\mathbf{\nabla^{2}} h(\mathbf{r})=&(AJ_{m}(\mu \rho) + BK_{m}(\mu \rho)) \times\nonumber \\&(C \cos m \theta + D \sin m \theta ) \approx AJ_{0}(\mu \rho) \label{eqLevSurfDeforSol}
\end{eqnarray}
where $J_{m}$ is the Bessel function, $K_{m}$ is the modified Bessel function, the values of $A$, $B$, $C$, $D$ are determined by the boundary conditions and the relation $\mu=\sqrt{(\sigma/k-\lambda^{2}/(k b))}$. In our physical case, it is sufficient to restrict the consideration to the most important solution where the amplitude $A$  may be found from the condition that the final curvature is equal to the maximum curvature of the micro-swimmer surface, $\mathbf{\nabla^{2}} h(\mathbf{r})=2/R^{2}=A $, where $R$ is the transverse radius of the micro-swimmer. We assume that the variables describing the state are interrelated and hence domains with different surfaces of local curvature are formed. Such deformation may vary along the surface and produce fluid flows responsible for the motion of the micro-swimmer.

We need to emphasize that Eq.~\ref{6} for the possible inhomogeneous deformation of the micro-swimmer's surface is obtained in the linear approximation. Its solution presented in Eq.~\ref{eqLevSurfDeforSol} just demonstrates the possibility of the formation of such deformations in the local, i.e., flat, geometry, that rapidly vanish and cannot provide micro-swimmer's steady motion. The mechanism of the nonlinear formation of a steady dipole concentration may be provided by the feedback of the concentration formed, which may be introduced in terms of the coefficient $b(I, c(r), T)$ that depends on the dipole concentration, and probable energy redistribution between individual components of the micro-swimmer. A one-dimensional nonlinear problem was considered in Refs.\cite{LZ, LTZ}, and its exact solutions are known. Our case concerns a three-dimensional problem that is much more complicated.

\subsection{Estimation of the velocity}
The theoretical non-uniform polarization distribution obtained above on the micro-swimmer's surface and correspondent deformation profile evolve along the surface. We consider the micro-swimmer's velocity provided by the mechanism under consideration proportional to the velocity of distribution or deformation evolution. However, before we present the numerical estimation of the micro-swimmer's velocity, we have to give the values of the quantities contained in the free energy functionals. So, the continuous mean-field order-disorder parameters $\alpha = \alpha (\rho) = \Delta_R - \gamma \rho / 2 \label{eqLevAlphaMarc} $, $ \beta = \beta (\rho) = \gamma^2 \rho^2 / 8 \Delta_R \label{eqLevBetaMarc}$ are taken from Ref.\cite{Mar}, while Frank constant is assumed to be $K \sim 10^{-10}~\mathrm{dyn}$. The rotary viscosity contained in the equations for $\alpha$ and $\beta$ is put to be $\gamma = 0.02 \mathrm{P}$,  while the translation correlation function for the noise, $\Delta_R = k_B T/\zeta_R = D_r$, is taken to be equal to the rotary diffusion coefficient because the system under consideration is assumed to be thermal and to satisfy the Stokes-Einstein relations for given temperature $T \approx 293~\mathrm{K}$ and rotary friction coefficient $\zeta_R = 8 \pi \eta R^3$ for a spherical cluster of radius $R \approx 2~\mathrm{cm}^{-4}$ with the friction coefficient $\eta = 0.05~\mathrm{P}$.

In view of the values given above, we propose the following method for the estimation of the micro-swimmer's velocity. It employs the dynamical equations for active soft matter contained \cite{Mar}, in particular, in the equation for the evolution of the vector polarization field, i.e.,
\begin{equation}
	\frac{\partial}{\partial t} \mathbf{p}\left( \mathbf{r}, t \right) + \lambda_1 \left( \mathbf{p} \cdot \boldsymbol{\nabla} \right) \mathbf{p} = - \frac{1}{ \gamma } \frac{ \delta F }{ \delta \mathbf{p} } + \mathbf{f}, \label{eqVecPolFieldEvol}
\end{equation}
In this equation, a specially highlighted left-hand side is reminiscent of the nonlinear Burgers equation, the soliton solutions of which have long been fully investigated with various nonlinearities in the free energy \cite{Whi}. For this reason, it seems reasonable to assume that such a soliton-like formation of an inhomogeneous polarization distribution can be realized in our case. Here we are interested mainly in the terms with the time derivative of the polarization vector $\mathbf{p}$ and the free energy functional $1/\gamma \times \delta F / \delta \mathbf{p} $, while the advection term and Gaussian noise term may be disregarded. We assume that the hydrodynamic flow is proportional to the velocity of the inhomogeneous polarization distribution, and therefore, a linear estimate of the velocity polarization profile can be proposed as \cite{jcp3158471, PhysRevE.56.3265} $V \sim K / \gamma l_{bact} \sim 10^{-4} \mathrm{cm/s}$.

We can also propose a few simple experimental verifications for the mechanism in addition to straightforward micro-swimmer's velocity dependence on luminosity intensity. The first one concerns the micro-swimmer's velocity dependence on the acidity, i.e., pH, of the aqueous medium since it involves the electric force. Another one considers the inhomogeneous distribution rotation frequency $\omega \ge \sqrt{\left( C \rho_{water} v^2_{bact} \right) / \left( \rho_{bact} r^2_{bact} \right)}$ that is governed by the balance between the kinetic rotation energy and medium resistance. Here, $C$ is the resistance coefficient for the micro-swimmer's shape.

\section{Discussion}
The motion mechanism presented above is constructed with applicability to living active matter systems. As an example of a physical system for this mechanism, we consider motile flagellaless bacteria, like flagellaless Chromatium okenii. There are ones with and without flagella \cite{DiNezio2024} among the bacterial population, and there are well-studied natural habitat conditions for these bacteria \cite{Sommer2017}. Moreover, this bacterium is considered to be responsible for mixing in natural waters \cite{Sommer2017} and the temperature, salinity and turbidity profiles in natural water support the assumption of free charges in medium needed for the proposed mechanism. Chromatium okenii is a single-cell, oblong, capable of anoxygenic photosynthesis, a phototactic bacterium of violet, red or green color and oval or drop-like shape with a flexible gel surface \cite{Luedin2019, DiNezio2024}. It is filled with moving sulfur globules, which are domains with higher densities and represent various dielectric clusters \cite{Hag}. It is a phototactic bacterium \cite{Luedin2019, DiNezio2024}, and its globules are optically active \cite{Hag}.

So, the assumption that luminous flux enhances polarization of globules \cite{Agar1989}, intensifies their natural continuous motion through the thermoelectric mechanism caused by additional spatially inhomogeneous temperature distribution between the illuminated and shaded bacterium parts in a similar way as for other optically active molecules \cite{Tabe2003, Rebesh2022, Iti, Kal, Uts, Mark, Bra} and provides them with torque looks natural. We suppose that globules have sufficient momentum to move inside the bacterium's volume. Since globule motion is governed by the inhomogeneous charge concentration and temperature distribution \cite{Iacopini2006, Alo, Pav, Wen, Kai, LAVRENTOVICH201697} the enhanced globule polarization with intensified globule motion and torque results in charge redistribution both within the bacterium's volume and on its surface that lead to the development of the non-uniform polarization distribution on the bacterium's surface, along with surface deformation and surface torque. We suppose that the globule's polarization, momentum and torque are all sufficient to non-uniformly redistribute the dipoles at the surface, soliton-like deform the surface itself and provide the surface with torque to evolve such soliton-like disturbance along the bacterium's surface in a form we obtained in section Results. We suppose that such disturbances of polarization and deformation at the bacterium's surface caused by self-consistent globule motion recur while an ambient light has sufficient intensity. The more accurate self-repetition of these disturbances and the higher frequency of them provide a higher velocity for the bacterium. Finally, we suppose that polarization magnitudes and surface deformation amplitudes are sufficient to provide a directional motion of the bacterium, in correspondence to the velocity estimation in the Results section. We suppose that the experimental results mentioned in the Introduction and Results, especially the manipulation of light effect by particles and generation of complex structures due to the orientation order, support the possibility of the functioning of individual elements that are assumed to occur in the motion mechanism proposed. Thorough analysis of these processes requires detailed knowledge of the intrinsic processes within the specific micro-swimmer, like the Chromatium okenii bacterium, allowance for exact geometry and variation of the environmental temperature and illumination intensity. These problems lie beyond the framework of this paper since its purpose is to demonstrate the possibility of the micro-swimmer's motion mechanism due to the interaction of the inhomogeneous polarization distribution on the micro-swimmer surface with ions contained in the water and to the interaction of the deformed micro-swimmer surface with water flows. We emphasize that this mechanism of micro-swimmer's motion is based on the induced self-consistent behavior of the active matter, in the case of Chromatium okenii, for example, polarized domains of sulfur globules. It is the process occurring within the micro-swimmer due to external factors. This feature makes it different from the known electrothermophoresis, which is an external process.

The mathematical description of the mechanism under consideration employs the thermodynamic approach in terms of the free energy functional. The equation for the surface deformation Eq.~\ref{eqLevSurfDeforSol} is solved in the linear approximation. The deformations in the linear approximation are shown to vanish fast and thus to be unable to provide a steady micro-swimmer's motion. Hence, the steady motion requires a mechanism or a nonlinear organization of a steady polarization inhomogeneity. Solutions of the equations for the non-equilibrium polarization distribution under small and strong perturbations of the polarization-dependent elasticity coefficient are shown in Fig.~\ref{fig_pKappaSmall:fig1}. The first one may be the reason for the micro-swimmer's motion, similarly to the polarization current accompanied by the modification of the local polarization of the medium, while the second one may be treated as a fragmentation of the micro-swimmer into two parts. An estimate of the velocity of a micro-swimmer is proposed in Eq.~\ref{eqVecPolFieldEvol} for such a motion mechanism under given natural conditions within the framework of known experimental observations, along with a few experimental verifications for the mechanism, like velocity dependence on luminosity intensity, medium's acidity and observable inhomogeneous distribution rotation frequency.

Finally, in laboratory experiments of domestication of Chromatium okenii, the bacterium's flagella become redundant \cite{DiNezio2024}, so such domesticated bacteria without flagella have lower motility in comparison with the flagellated non-domesticated ones \cite{DiNezio2024}. In this study, this fact is presented as "low/to no" motility with velocities equal to or less than $5~\mu m /s$, which corresponds to the estimation by magnitude we provided for the proposed mechanism in section Results. It is very important to underline that such velocities is almost by magnitude higher than the velocity of motile Myxococcus xanthus with a gliding motion mechanism \cite{Chen2023} and similar to the flagellaless Spiroplasma one. Taking into account all the above, we consider the proposed mechanism as a plausible candidate for the flagellaless Chromatium okenii motion mechanism, whose motility cannot be explained by simple advection, thermoelectrophoresis \cite{LAVRENTOVICH201697, Turiv2020} or bioconvection processes. This result motivates further experimental studies and more exact velocity calculations.

\section{Conclusion}
We proposed the new mechanism for micro-swimmer motion in an aquatic environment, which implies light-induced polarization perturbations as a motivation for its motion. This mechanism assumes that natural light leads to an enhancement of the non-uniform polarization distribution at the micro-swimmer's surface and also its surface deformations. The interaction of such non-uniform polarization with free charges of the medium and the mechanical interaction of surface deformations with media flows lead to the directed micro-swimmer motion. For the oblong drop-like shaped micro-swimmer with elastic surface, the free energy functional with polarization-dependent elasticity coefficient is minimized, and corresponding non-uniform polarization distributions are obtained. The corresponding surface deformations are presented in linear approximation. The obtained solutions correspond to the motion along the micro-swimmer or the rotation of the non-uniform polarization distribution. The estimation of the velocity of micro-swimmer motion, which is based on the hydrodynamics of the active matter, is obtained and corresponds in magnitude to the experimentally observed values. The experimental verifications for the mechanism through velocity dependence on luminosity intensity, medium's acidity and observable inhomogeneous distribution rotation frequency are noted.

\backmatter

\bmhead{Acknowledgements}

B.L. is grateful for the support of the National Research Foundation of Ukraine through the project 2023.03/0165 "Quantum Correlations of Electromagnetic Radiation", during the implementation of which this work was actually carried out, although the main idea arose earlier during participation in the grant P1-0099 of the Slovenian Research Agency (ARRS).

O.C. is grateful for the support of the Simons Foundation through the project "Stochastic Processes in Condensed Media, Biological Systems and Radiation Fields" 0125U000031.

Both authors are also sincerely grateful to Miha Ravnik and Volodymyr Zasenko for insightful comments and thorough discussions of the ideas presented in this article.

\section*{Conflicts of interest}
There are no conflicts to declare

\section*{Data availability}
All data generated or analysed during this study are included in this published article.

\section*{Author contributions}
B.L. formulated the problem, gave the idea of the mechanism and presented equations, O.Ch. solved equations, gave interpretation of results and wrote text. All authors reviewed the manuscript.

\bibliography{pmbmamRef_epjp}

\end{document}